\documentclass[twocolumn]{aastex62}
  
\usepackage[utf8]{inputenc}
\usepackage{comment}
\usepackage{subfigure}
\usepackage{amsmath}
\usepackage{amssymb}
\usepackage{hyperref}
\usepackage{bm}
\usepackage{xcolor}
\usepackage{acronym}   
\usepackage{units}
\usepackage{xspace}
\usepackage{tabularx}
\usepackage{float}
\usepackage{pifont}
\usepackage{ifthen}

\definecolor{ochre}{rgb}{0.8, 0.47, 0.13}

\ifthenelse{1>0}
{
\newcommand{\reinhold}[1]{\textcolor{green}{\textbf{RW: #1}}}
\newcommand{\ilya}[1]{\textcolor{magenta}{\textbf{Ilya: #1}}}
\newcommand{\eric}[1]{\textcolor{blue}{\textbf{Eric: #1}}}
\newcommand{\simon}[1]{\textcolor{orange}{\textbf{Simon: #1}}}
\newcommand{\avg}[1]{{\color{ochre}\textbf{AVG: #1}}}
\newcommand{\adam}[1]{\textcolor{red}{\textbf{Adam: #1}}}
\newcommand{\resp}[1]{{\color{red}\textbf{#1}}}
}{
\newcommand{\reinhold}[1]{}  
\newcommand{\ilya}[1]{}  
\newcommand{\eric}[1]{}  
\newcommand{\simon}[1]{}  
\newcommand{\avg}[1]{}  
\newcommand{\adam}[1]{} 
\newcommand{\resp}[1]{}
}

\newcommand{\Msun}{\ensuremath{\xspace\rm{M}_{\odot}}\xspace}
\newcommand{\Zsun}{\ensuremath{\xspace\rm{Z}_{\odot}}\xspace}
\newcommand{\kms}{\ensuremath{\xspace\rm{km}\,\rm{s}^{-1}}\xspace}
\newcommand{\flow}{\ensuremath{f_\textrm{low}}\xspace}

\newcommand{\vcut}{\ensuremath{v_\textrm{cut}}\xspace}
\newcommand{\Nslow}{\ensuremath{N_\textrm{slow}}\xspace}
\newcommand{\Np}{\ensuremath{N_P}\xspace}
\newcommand{\Pflow}{\ensuremath{{P({\leq} \Nslow | \flow)}}\xspace}

\newcommand{\twoSigma}{\ensuremath{2\sigma}\xspace}
\newcommand{\posTwoSigma}{\ensuremath{+2\sigma}\xspace}
\newcommand{\negTwoSigma}{\ensuremath{-2\sigma}\xspace}
\newcommand{\pmTwoSigma}{\ensuremath{\pm2\sigma}\xspace}
\newcommand{\xmark}{\textcolor{red}{\ding{55}}}
\newcommand{\cmark}{\textcolor{olive}{\checkmark}}

\newcommand{\fiducial}{\textsc{Fid}\xspace}
\newcommand{\kickOne}{\textsc{K1}\xspace} 
\newcommand{\noWideECSNe}{\textsc{NW}\xspace}
\newcommand{\noWideECSNeKickOne}{\textsc{NW-K1}\xspace}
\newcommand{\longfiducial}{\textsc{Fiducial}\xspace}
\newcommand{\longkickOne}{\textsc{Kick: \unit[1]{\kms}}\xspace} 
\newcommand{\longnoWideECSNe}{\textsc{No Wide ECSNe}\xspace}
\newcommand{\longnoWideECSNeKickOne}{\textsc{No Wide ECSNe, Kick: \unit[1]{\kms}}\xspace}

\acrodef{ZAMS}{zero-age main sequence}
\acrodef{NS}{neutron star}
\acrodef{BH}{black hole}
\acrodef{WD}{white dwarf}
\acrodef{CO}{compact object}
\acrodef{MSP}{millisecond pulsar}
\acrodef{CDF}{cumulative distribution function}
\acrodef{BBH}{binary black hole}
\acrodef{DNS}{double neutron star}
\acrodef{DCO}{double compact object}
\acrodef{sAGB}{super-Asymptotic Giant Branch}

\acrodef{GC}{globular cluster}
\acrodef{DM}{dispersion measure}
\acrodef{VLBI}{very long baseline interferometry}
\acrodef{TOA}{time of arrival}
\acrodef{LSR}{local standard of rest}
\acrodef{CDF}{cumulative distribution function}
\acrodefplural{CDF}[CDFs]{cumulative distribution functions}

\acrodef{RLOF}{Roche-lobe overflow}
\acrodef{CE}{common envelope}
\acrodef{SN}{supernova}
\acrodefplural{SN}[SNe]{supernovae}
\acrodef{ECSN}{electron-capture supernova}
\acrodefplural{ECSN}[ECSNe]{electron-capture supernovae}
\acrodef{USSN}{ultra-stripped supernova}
\acrodefplural{USSN}[USSNe]{ultra-stripped supernovae}
\acrodef{CCSN}{core-collapse supernova}
\acrodefplural{CCSN}[CCSNe]{core-collapse supernovae}

\newcommand{\CCps}{\mbox{CC-pulsars}\xspace}

\newcommand{\ECps}{\mbox{EC-pulsars}\xspace}

\acrodef{COMPAS}{Compact Object Mergers: Population Astrophysics and Statistics}
\acrodef{BPS}{binary population synthesis} 
\acrodef{SSE}{single star evolution} 
\acrodef{BSE}{binary star evolution} 
\acrodef{IMF}{initial mass function}

\shorttitle{Weak supernova kicks}
\shortauthors{Willcox et al.}

\begin{document}

\title{Constraints on Weak Supernova Kicks from Observed Pulsar Velocities}

\correspondingauthor{Reinhold Willcox}
\email{reinhold.willcox@monash.edu}

\author[0000-0003-0674-9453]{Reinhold Willcox}
\affiliation{School of Physics and Astronomy
Monash University,
Clayton, VIC 3800, Australia}
\affiliation{
The ARC Centre of Excellence for Gravitational Wave Discovery -- OzGrav, Australia}
\author[0000-0002-6134-8946]{Ilya Mandel}
\affiliation{School of Physics and Astronomy
Monash University,
Clayton, VIC 3800, Australia}
\affiliation{
The ARC Centre of Excellence for Gravitational Wave Discovery -- OzGrav, Australia}
\affiliation{Institute of Gravitational Wave Astronomy and School of Physics and Astronomy, University of Birmingham, Birmingham, B15 2TT, United Kingdom}
\author[0000-0002-4418-3895]{Eric Thrane}
\affiliation{School of Physics and Astronomy
Monash University,
Clayton, VIC 3800, Australia}
\affiliation{
The ARC Centre of Excellence for Gravitational Wave Discovery -- OzGrav, Australia}
\author[0000-0001-9434-3837]{Adam Deller}
\affiliation{Centre for Astrophysics and Supercomputing, Swinburne University of Technology, Hawthorn, VIC 3122, Australia}
\affiliation{
The ARC Centre of Excellence for Gravitational Wave Discovery -- OzGrav, Australia}
\author[0000-0002-6100-537X]{Simon Stevenson}
\affiliation{Centre for Astrophysics and Supercomputing, Swinburne University of Technology, Hawthorn, VIC 3122, Australia}
\affiliation{
The ARC Centre of Excellence for Gravitational Wave Discovery -- OzGrav, Australia}
\author[0000-0003-1817-3586]{Alejandro Vigna-G\'omez}
\affiliation{
DARK, Niels Bohr Institute,
University of Copenhagen,
Jagtvej 128, 2200,
Copenhagen, Denmark
}


\begin{abstract}

Observations of binary pulsars and pulsars in globular clusters suggest that at least some pulsars must receive weak natal kicks at birth.  If all pulsars received strong natal kicks above  \unit[50]{\kms}, those born in globular clusters would predominantly escape, while wide binaries would be disrupted.  On the other hand, observations of transverse velocities of isolated radio pulsars indicate that only $5\pm2\%$ have velocities below \unit[50]{\kms}.  We explore this apparent tension with rapid binary population synthesis modelling.  We propose a model in which supernovae with characteristically low natal kicks (e.g., electron-capture supernovae) only occur if the progenitor star has been stripped via binary interaction with a companion.  We show that this model naturally reproduces the observed pulsar speed distribution and without reducing the predicted merging double neutron star yield.  We estimate that the zero-age main sequence mass range for non-interacting progenitors of electron-capture supernovae should be no wider than ${\approx}\unit[0.2]{\Msun}$.

\end{abstract}



\section{Introduction} \label{sec:intro}

The observed single pulsar population is characterized by typical speeds of hundreds of \kms, generally attributed to large natal kicks due to asymmetric mass ejection during \acp{SN} \citep{Hobbs05, Burrows13, Verbunt17, Deller19}. Such large kicks are consistent with simulations of \acp{CCSN} \citep{Wongwathanarat12, Muller20}.  On the other hand, the existence of pulsars in globular clusters, where escape velocities can be ${\lesssim}\unit[50]{\kms}$, points to the need for a subpopulation of pulsars with low kicks \citep[e.g.,][]{Sigurdsson:2003}.  Meanwhile, natal kicks that disrupt the binary inhibit the formation of \acp{DNS} \citep[e.g.,][]{BeniaminiPiran:2016, VignaGomez18}.  Therefore, at least some fraction of \acp{NS} must receive low natal kicks.

Theoretically, low kicks (${\lesssim}\unit[30]{\kms}$) are often associated to \acp{ECSN} and \acp{USSN}. \acp{USSN} only occur as the second \ac{SN} in very tight binaries with minimal mass loss and almost never disrupt the binary; thus, we do not consider them further as they do not contribute to the velocity distribution of isolated pulsars \citep{Tauris:2015}.  \acp{ECSN} are thought to arise from a subset of the \ac{sAGB} stars, which span a \ac{ZAMS} mass range of ${\approx}\unit[6.5-12]{\Msun}$, though the precise \ac{ZAMS} mass range for \ac{ECSN} progenitors is highly uncertain (see \citealt{Doherty17} for a review).

In this paper, we use the \href{https://compas.science/}{COMPAS} rapid binary population synthesis code \citep{Stevenson:2017, VignaGomez18} in order to reconcile the paucity of observed low-velocity pulsars with the need for a population of \acp{NS} with low natal kicks.  We find that by restricting low natal kicks to occur only in stars which have previously transferred mass to a binary companion, our model reproduces the observed fraction of low-velocity isolated pulsars, without inhibiting the \ac{DNS} yield.  We interpret this as an upper limit on the \ac{ZAMS} mass range for effectively single progenitors of \acp{ECSN}, while leaving the channel for stripped stars unaffected.  

Our proposal builds on indications that \acp{ECSN} from single stars are less common than initially thought \citep{Miyaji80, Nomoto:1984, Poelarends:2008}, but could be enhanced in binaries because envelope stripping by Roche-lobe overflow onto a companion suppresses second dredge-up \citep{Podsiadlowski04, VanDenHeuvel10, Ibeling13, DallOsso14, Poelarends:2017}. 

In Sec.~\ref{sec:data}, we discuss the data set of pulsar velocities observed with very long baseline interferometry and the associated selection effects.  In Sec.~\ref{sec:methods}, we describe our population synthesis prescriptions.  We present the results in  Sec.~\ref{sec:results}.  In Sec.~\ref{sec:discussion}, we discuss the caveats of our analysis and its implications for stellar evolution modellers.


\section{Data and Selection Effects}
\label{sec:data}

The data used in this study are astrometric measurements of isolated pulsars obtained with very long baseline interferometry (VLBI).  We use exclusively VLBI measurements rather than the larger data sets based on pulsar timing or dispersion measure because VLBI provides precise measurements and we aim to avoid concerns about systematic uncertainty in other pulsar velocity measurements \citep{Deller19}.

We select a total of 81 pulsars from a variety of studies, primarily \citet{Deller19} but also \citet{Bailes1990b, Fomalont99, Chatterjee01,  Brisken02, Brisken03b, Dodson03, Chatterjee04, Chatterjee09, Deller09} and \citet{Kirsten15}.  We remove millisecond pulsars, pulsars known to be in binaries, and pulsars in globular clusters in order to obtain a relatively homogeneous data set of pulsars whose velocities are primarily set by \ac{SN} natal kicks (and any previous isolated binary evolution) rather than dynamics.  Some of the apparently single pulsars may still be in binaries, as binaries with orbital periods much longer than the observational baseline are challenging to detect.

The data for each pulsar contain sets of bootstrapped fits for the parallax, position, and proper motion.  We treat the fits as posterior samples.  We re-weigh them by applying a prior on the parallax to be above 0.05 mas, i.e., the distance to be below 20 kpc, consistent with Galactic pulsars, as well as a prior on the transverse velocities to be less than \unit[2000]{\kms}, to exclude anomalously large values.  

We extract an intrinsic transverse velocity relative to the pulsar's \ac{LSR} by correcting for Galactic rotation and the motion of the Sun.  We assume a flat Galactic rotation curve, with constant speed \unit[230]{\kms} \citep{Bhattacharjee14}, and take solar velocity (U, V, W)$_\odot$ = \unit[(11.1, 244, 7.25)]{\kms} \citep{Schonrich10}.  This correction assumes that the \ac{LSR} velocity vector of the pulsar today is similar to its \ac{LSR} at birth, which relies on the assumption of insignificant acceleration in the Galactic potential.  This is valid for very young or slow-moving pulsars, but is not justified, e.g., for a \unit[20]{Myr} pulsar moving at \unit[500]{\kms}, which would have travelled for ${\sim}\unit[10]{kpc}$ in the absence of external forces.  However, only pulsars with high velocities (comparable to or larger than the local Galactic rotational velocity) will be impacted, so this is not expected to affect our analysis of low-velocity pulsars.  A bigger concern is that the data were not collected in a manner intended to be a complete survey.  Indeed, \citet{Deller19} note that their sample focused specifically on systems at high Galactic latitudes, which could lead to a paucity of low-velocity pulsars.  However, the distributions of characteristic ages and heights off the Galactic plane are indistinguishable between our sample and the broader sample of all isolated, non-millisecond pulsars in the ATNF pulsar catalogue\footnote{ \url{https://www.atnf.csiro.au/research/pulsar/psrcat/}}.

\begin{figure}[tpb]
\centering
\includegraphics[width=\columnwidth]{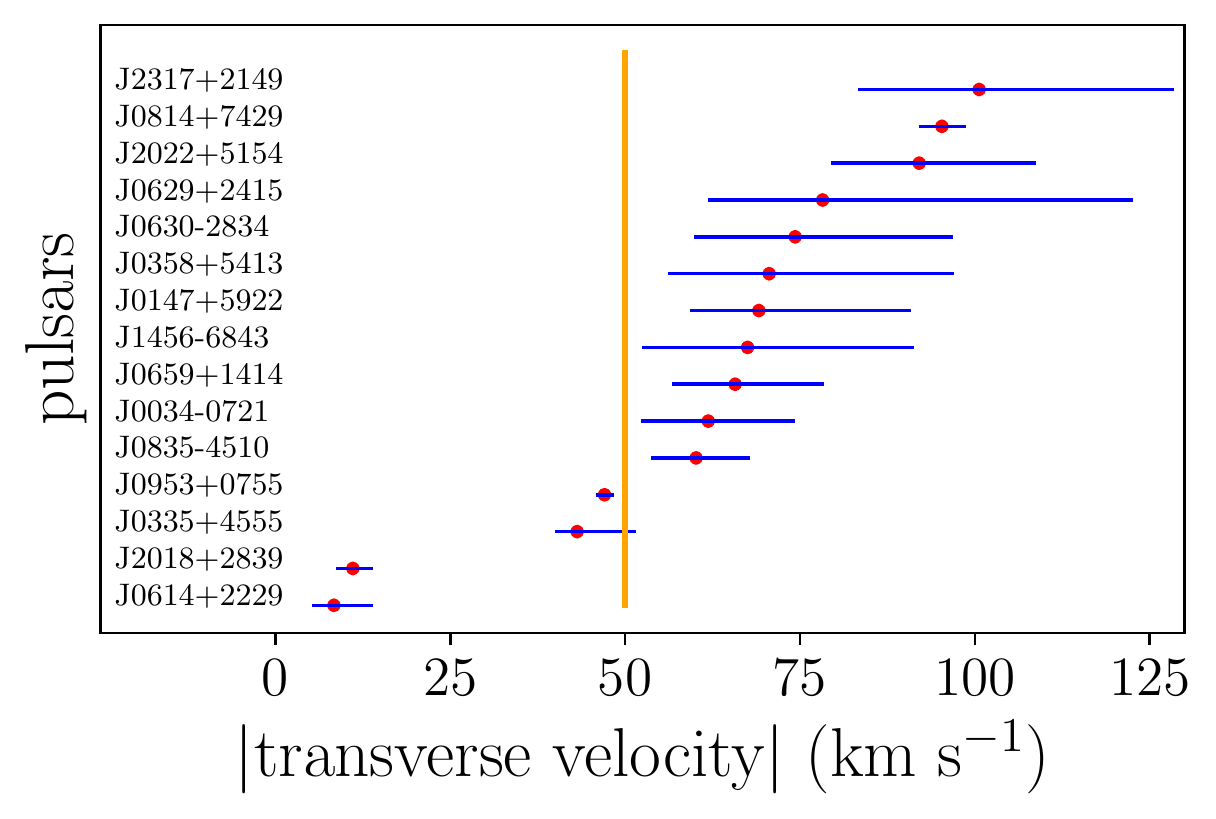}
\caption{
Transverse velocity medians and 5-95\% confidence intervals for the 15 lowest-velocity pulsars in our sample.  Defining \unit[50]{\kms} as the boundary between low- and high-speed pulsars is convenient since no pulsar shows strong support on both sides of this value.
}
\label{fig:vTrans}
\end{figure}

It is also conceivable that pulsars with low velocities are naturally fainter and harder to detect because of some correlation between their formation physics and an observational characteristic relevant to radio detectability that is not yet understood \citep{Sigurdsson:2003}.  We test for this possible bias by examining the correlation between pulsar transverse velocities and distances.  A Malmquist bias favours detecting only radio-bright pulsars at greater distances.  Therefore, if low-velocity pulsars tend to be fainter in the radio band, we should see a positive correlation between distance and transverse velocity.  While there is a mild correlation between distance and velocity, we confirmed that its magnitude is fully consistent with arising from the construction of transverse velocity as a product of proper motion and distance, with significant uncertainties in the distance (typically of order 20\%).  We therefore see no evidence for a selection effect against low-velocity pulsars once this correlation is accounted for. Further, we find that the correlation between the transverse velocities and heights of the pulsars out of the Galactic plane is no greater than the correlation between the transverse velocities and pulsar distances, which further supports the absence of a significant bias due to preferentially observing pulsars at high Galactic latitudes. Fig.~\ref{fig:vTrans} shows the transverse velocity distributions at the low end of our catalog.  


\section{Methods} \label{sec:methods}

We use the COMPAS rapid binary population synthesis code \citep{Compas21} to generate synthetic pulsar transverse velocity distributions for several different \ac{SN} kick prescriptions.  We simulate a total of $10^6$ binaries per prescription, with primary \ac{ZAMS} mass drawn from the \citet{Kroupa01} initial mass function between 5 and \unit[150]{\Msun}, mass ratio drawn uniformly from 0.01 to 1, and semi-major axis distributed uniformly in the log between 0.01 and \unit[1000]{AU}  \citep{Opik:1924}.  Here, very wide, non-interacting binaries represent the single star population.  All binaries are initially circularized ($e_0 = 0$) and use solar metallicity $\Zsun = 0.0142$ \citep{Asplund:2009}.  Unless otherwise specified, we follow the default COMPAS prescriptions \citep{Stevenson:2017, VignaGomez18, Vinciguerra20,Compas21}.  

We assume that a star with a helium core mass above \unit[2.25]{\Msun} at the base of the asymptotic giant branch will undergo \ac{CCSN} once the mass of its carbon-oxygen core reaches the threshold set in Eq.~(75) of \citet{Hurley00}, in which we replace the Chandrasekhar mass with \unit[1.38]{\Msun} \citep{Belczynski08}.   We set the \ac{CCSN} remnant mass according to the \citet{Fryer12} delayed prescription.  If the helium core mass is between 1.6 and $\unit[2.25]{\Msun}$ at the base of the asymptotic giant branch and the carbon-oxygen core mass subsequently reaches $\unit[1.38]{\Msun}$, the star is assumed to form a \ac{NS} with mass $\unit[1.26]{\Msun}$ in an \ac{ECSN}.

We consider a variety of different kick prescriptions, all based on the following ``fiducial'' prescription.  Natal kicks for pulsars formed in \acp{CCSN} (hereafter, \CCps, and similarly for \ECps) are drawn directly from the observed pulsar velocity distribution.  To obtain a 3D kick magnitude for the \CCps, we randomly draw a velocity sample from the union of the transverse velocity posterior distributions of pulsars in our catalogue, selecting only among values of at least \unit[50]{\kms}, and divide by a random projection coefficient under the assumption of an isotropic viewing angle.

This is in contrast to the COMPAS default \ac{CCSN} kick prescription (which uses the the \citealt{Hobbs05} Maxwellian model) and other evolution codes which draw from parametrized, analytical distributions, because these distributions differ from the observations at high velocities.  However, this difference is due directly to the assumed shape of the natal kick distribution.  The bulk of the isolated pulsar population have very high speeds ${\gtrsim}\unit[300]{\kms}$, much larger than typical pre-\ac{SN} orbital velocities, 
\begin{equation*}
v_\text{orb} \approx \unit[42]{\kms} \left( \frac{M}{\unit[20]{\Msun}} \right)^{1/2} \left( \frac{a}{\unit[10]{\text{AU}}} \right)^{-1/2}.
\end{equation*}
where $M$ is the total binary mass and $a$ is the separation of a circular binary.  Consequently, the pulsar is nearly always unbound, with a final speed dominated by the magnitude of the natal kick, with some scatter on the order of $v_\text{orb}$.  By drawing \ac{CCSN} kicks directly from the observations, we are able to focus on the poorly-understood low-velocity regime. 

In the \longfiducial prescription (henceforth abbreviated as \fiducial), all \ECps have natal kick magnitudes drawn from a Maxwellian with $\sigma_\text{1D rms} = \unit[30]{\kms}$.

All \ac{SN} natal kicks are applied isotropically in the progenitor rest frame (cf. \citealt{BrayEldridge18} and \citealt{GiacobboMapelli20}, who investigate anisotropic kicks).  If the \ac{SN} disrupts the binary through a combination of symmetric mass loss (the Blaauw kick; \citealp{Blaauw61}) and the natal kick, the speed of the \ac{NS} is given by its asymptotic speed once it has escaped the gravitational potential of the companion.  If the binary remains intact, the speed of the \ac{NS} is given by the sum in quadrature of its speed in the center-of-mass frame and the binary velocity relative to the \ac{LSR}, if any. We then project the \ac{NS} speed onto the plane of the sky assuming an isotropically distributed viewing angle to obtain a transverse velocity prediction.  \acp{NS} in wide binaries with orbital periods above 10 years are categorized together with isolated \acp{NS} as ``apparently isolated''; varying the period threshold between 10 and 100 years has a negligible impact.  Wide binary pulsars comprise ${\sim} 20\%$ of the total apparently isolated \ac{NS} population, immediately following the \ac{SN}, in all our prescriptions.   This fraction is a few times higher than the estimated fraction of pulsars in wide binaries obtained by \citet{Antoniadis:2021} ($\lesssim 10\%$), who considered observability through Gaia and radio pulsar surveys under a very simple model of binary properties (see also \citealt{IgoshevPerets:2019}).  This discrepancy is in part due to the fact that our 20\% estimate does not account for the amount of time that a pulsar is observable in a wide binary, which could be disrupted by a second supernova.  It would be interesting to compare the properties of wide binaries predicted by our models against observations through a similar modelling of selection effects.

\acp{ECSN} are the primary source of low-velocity pulsars in our models, so we consider variations on the \fiducial prescription described above that impact \acp{ECSN}.  In the first variation, \longkickOne (\kickOne), we reduce the natal kicks of \acp{ECSN} down to a Maxwellian with $\sigma_\text{1D rms} = \unit[1]{\kms}$, motivated in part by hydrodynamical simulations \citep{GessnerJanka18}. The second variation, \longnoWideECSNe (\noWideECSNe), flags and ignores \acp{ECSN} in non-interacting binaries, allowing for \acp{ECSN} only in stars which previously lost mass through Roche-lobe overflow.  The final variation, \longnoWideECSNeKickOne (\noWideECSNeKickOne), combines the first and second variations.  A list of variations is given in Table~\ref{tab:stats}.


\section{Results} \label{sec:results}

\begin{figure}
\centering
\subfigure[Fiducial (\fiducial)]{
  \includegraphics[width=\columnwidth]{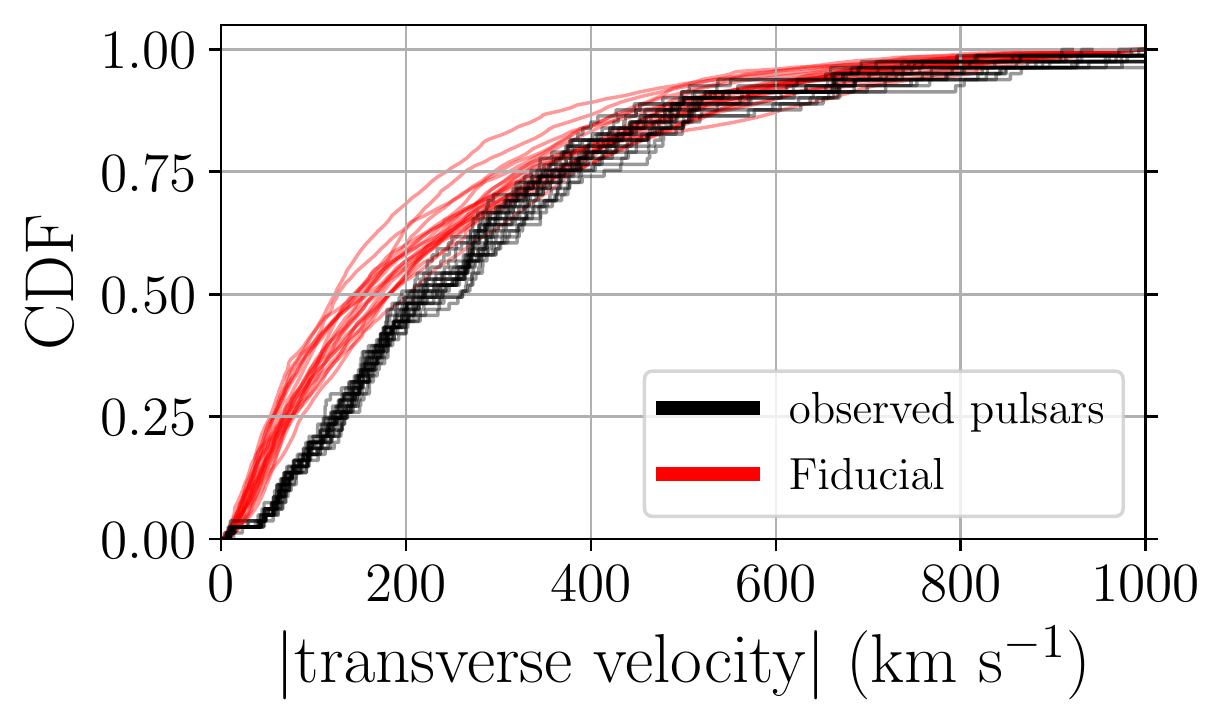}
}
\subfigure[No Wide ECSNe (\noWideECSNe)]{
  \includegraphics[width=\columnwidth]{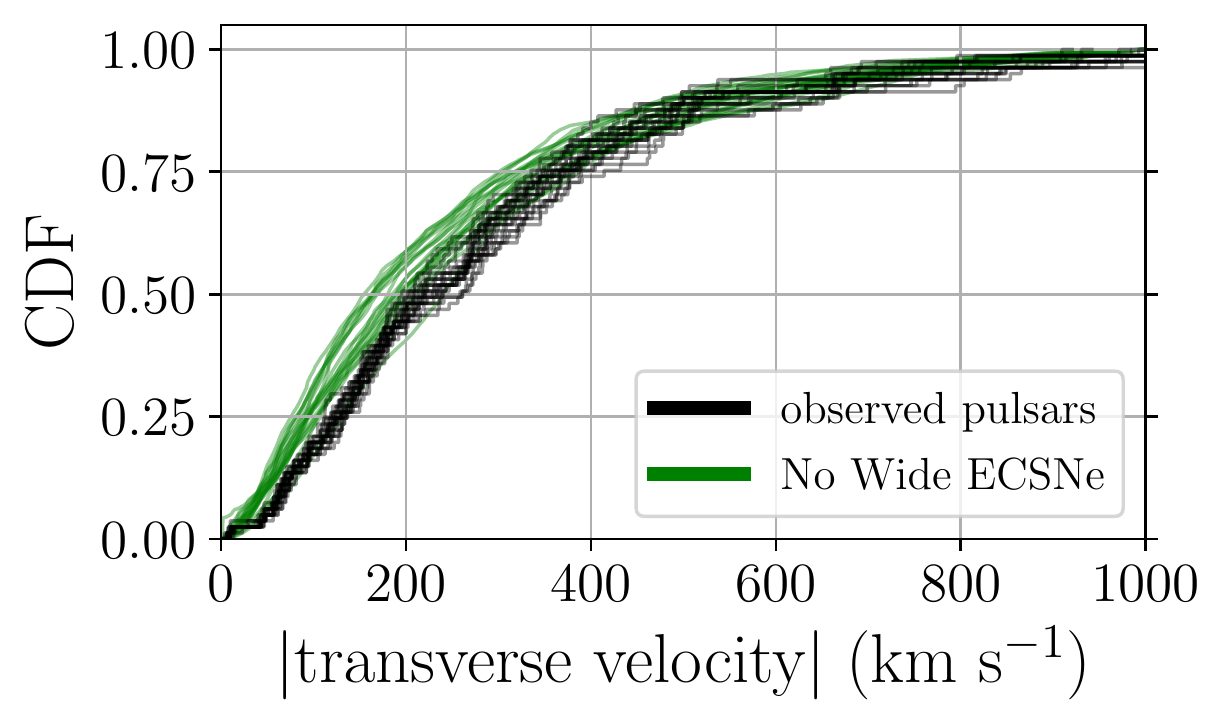}
}
\caption{
CDFs of the observed pulsar transverse velocities (black curves) and those predicted by the model variations (colored curves).
Families of CDFs are drawn to illustrate uncertainty (see Sec.~\ref{sec:results}). 
The \fiducial variation (a) over-produces low-speed pulsars ($\lesssim\unit[50]{\kms}$).
The \noWideECSNe variation (b) improves the match at low speeds by removing \ac{ECSN} progenitors that did not transfer mass onto a binary companion.
}
\label{fig:cdfs}
\end{figure}

In Fig.~\ref{fig:cdfs}, we compare the transverse velocity \acp{CDF} of apparently isolated \acp{NS} from the \fiducial (a) and \noWideECSNe (b) prescriptions to those of the observed pulsars.  Each observed transverse velocity \ac{CDF} (black) is constructed by randomly sampling one posterior transverse velocity sample per pulsar.  The spread in these \acp{CDF} indicates the uncertainty in the velocity measurements.  Each synthetic \ac{CDF} (colored) is constructed by randomly drawing as many  velocities as the total number of observed pulsars $\Np{=}81$ from the modelled population.  Their spread indicates the impact of small-number statistics.  The \acp{CDF} match at velocities ${\gtrsim}\unit[400]{\kms}$ (both panels), validating the use of the \ac{CCSN} prescription used in this study. 

Meanwhile, the mismatch between the modelled and observed velocity distributions in (a) at low velocities indicates that the \fiducial prescription overpredicts the number of low-velocity pulsars.  The preferred \noWideECSNe prescription (b) reduces the disagreement in the low-velocity regime by removing \acp{NS} from \acp{ECSN} in non-interacting binaries.

We devise a statistic in order to quantify the goodness of fit between model and data.  For a specified cutoff velocity \vcut, let \flow represent the fraction of low-speed pulsars with transverse velocity $\leq$ \vcut, as predicted by a given model variation.  We choose a value of $\vcut=\unit[50]{\kms}$.  While this value is somewhat ad hoc, it serves to differentiate between high kicks, which disrupt binaries and eject pulsars from globular clusters, and low kicks, which do not.  Let \Nslow be the number of observed pulsars with velocity $\leq$ \vcut out of \Np.  The probability of observing \Nslow low-velocity pulsars out of \Np is described by the binomial distribution.  The probability of observing \Nslow pulsars \emph{or fewer} is thus given by
\begin{equation}
  \Pflow =  \sum_{i=0}^{\Nslow} C_i^{\Np} (\flow)^i (1-\flow)^{\Np-i}.
\end{equation}
In practice, \Nslow is not known precisely due to the measurement uncertainty in observed pulsar transverse velocities, so it is marginalized out.  Under the null hypothesis that the data are drawn from the model variation, \Pflow is uniformly distributed, allowing us to rule out any variations which yield very small or very large values for \Pflow.

\begin{figure}
\centering
\subfigure[Including slow pulsars from CCSNe]{
  \includegraphics[width=\columnwidth]{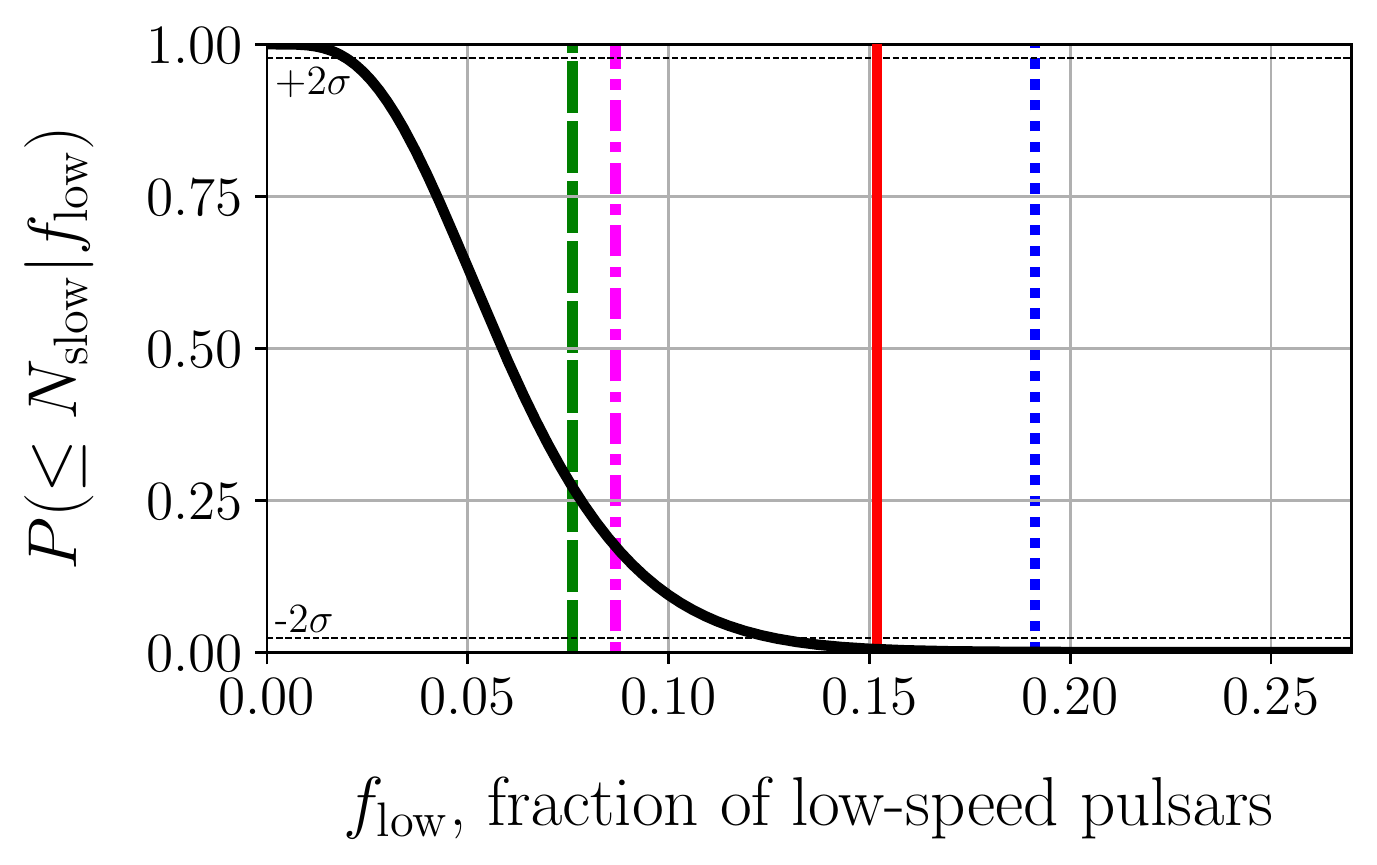}
}

\subfigure[Assuming fast kicks from CCSNe]{
  \includegraphics[width=\columnwidth]{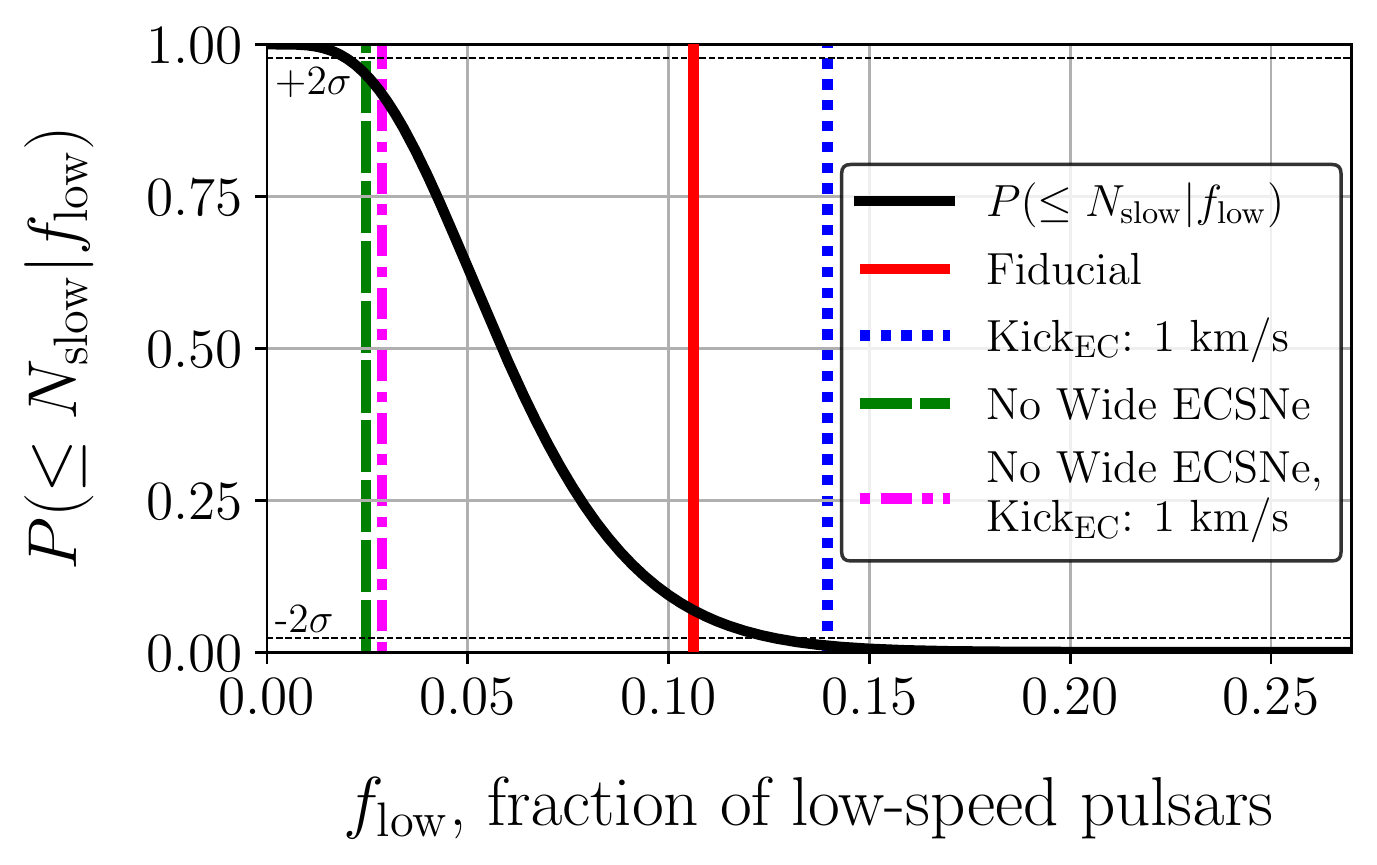}
}

\caption{
Probability \Pflow (black curve) that the observations would contain no more than the observed number \Nslow of low-speed (${\leq}\unit[50]{\kms}$) pulsars out of $\Np{=}81$, given a true fraction \flow of slow pulsars (see text).
Also plotted are model predictions for the slow pulsar fraction \flow (colored, vertical lines).
In (a), \flow includes slow pulsars from both \acp{CCSN} and \acp{ECSN}.  
In (b), \flow includes only slow \ECps.
Models that cross the black curve outside the \pmTwoSigma confidence bounds are in tension with observations (though see text for caveats).
}
\label{fig:pflow}
\end{figure}

\Pflow is plotted in Fig.~\ref{fig:pflow} in black.  The simulated \flow values are shown as colored vertical lines; the uncertainty due to the finite simulated population is within the width of the lines.  The null hypothesis is in tension with the data when a model, specified by \flow, intersects the black curve outside the indicated \pmTwoSigma interval (see Table~\ref{tab:stats} for exact values of \Pflow).

Fig.~\ref{fig:pflow}(a) shows \flow values calculated after including low-velocity pulsars from both \acp{CCSN} and \acp{ECSN}.  Model variations with \flow values which cross the \Pflow curve below the \negTwoSigma threshold over-produce low-velocity pulsars from a combination of \acp{ECSN} and \acp{CCSN}.  This is the case for the \fiducial and \kickOne variations.  Meanwhile, both variations which mask out \acp{ECSN} in non-interacting binaries cross well above the threshold and cannot be ruled out.

In order to distinguish the relative importance of \acp{ECSN} kicks and low-velocity \acp{CCSN} kicks, in Fig.~\ref{fig:pflow}(b), we consider an alternative population in which we include only \ECps as low-velocity pulsars.  Explicitly, we assume that all \CCps are given very large kicks.  Now, only the \kickOne model still crosses below the \negTwoSigma cutoff, so it is the only model that can be confidently ruled out on the basis that it over-produces low-velocity pulsars from \acp{ECSN} alone.  The low-velocity \ECps in this model predominately come from binaries disrupted by  Blaauw kicks, though ${\sim}30\%$ are in intact, very wide binaries.



The \fiducial model now falls just within the \twoSigma confidence interval and cannot be confidently ruled out.  However, the high production of low-velocity \ECps alone indicates a likely tension with observations.  The two models which mask non-interacting \acp{ECSN}, \noWideECSNe and \noWideECSNeKickOne, now nearly cross above the \posTwoSigma level, suggesting that they significantly \emph{under-produce} low-velocity pulsars.  Since we have intentionally removed \CCps here, this merely suggests that some fraction of low-velocity pulsars may be contributed by \CCps.

\begin{table*}
\centering
\begin{tabular}
{|>{\raggedright\arraybackslash}p{.20\textwidth}|l|l|l|>{\raggedright\arraybackslash}p{.2\textwidth}|}
\toprule
\multicolumn{1}{|l|}{Model} & 
\multicolumn{1}{p{.16\textwidth}|}{\Pflow including slow pulsars from \acp{CCSN}} & 
\multicolumn{1}{p{.16\textwidth}|}{\Pflow excluding slow pulsars from \acp{CCSN}} & 
\multicolumn{1}{p{.14\textwidth}|}{Merging DNS yield per $10^6$ $M_\odot$} & 
\multicolumn{1}{l|}{Variation} \\
\hline
\longfiducial(\fiducial) & 0.0049 \xmark & 0.069 \cmark & 7.1 $\pm$ 0.3 & --- \\
\longkickOne(\kickOne) & 0.0003 \xmark & 0.011 \xmark & 8.1 $\pm$ 0.3 & All ECSN kicks are reduced to \unit[1]{\kms} \\
\longnoWideECSNe(\noWideECSNe) & 0.2716 \cmark & 0.950 \cmark & 7.1 $\pm$ 0.3 & Non-interacting ECSN progenitors are removed \\
\longnoWideECSNeKickOne(\noWideECSNeKickOne) & 0.1741 \cmark & 0.918 \cmark & 7.5 $\pm$ 0.3 & Non-interacting ECSN progenitors are removed, other ECSN kicks are reduced to \unit[1]{\kms} \\
\hline
\end{tabular}
\caption{
List of model names and descriptions considered in this study, along with the \Pflow value derived if we include and exclude \CCps, and the yield of \acp{DNS} merging in a Hubble-time per \unit[$10^6$]{\Msun} of star formation.  Models with $\Pflow < 0.023$ over-produce slow pulsars at a rate inconsistent with observations at a $2\sigma$ level, as indicated by the check and cross marks.
}
\label{tab:stats}
\end{table*}


\section{Discussion}\label{sec:discussion}

We model supernova progenitors using binary population synthesis in order to self-consistently test \ac{NS} natal kick models against observed pulsar transverse velocities.  We find that our \fiducial variation over-produces low-speed, apparently isolated pulsars in comparison with observations.  This discrepancy can be resolved if supernovae which produce low-speed, apparently isolated \acp{NS} are suppressed in wide binaries when the \ac{NS} progenitor did not experience Roche-lobe overflow onto a companion.

If \acp{ECSN} indeed produce weak natal kicks, this provides a constraint for models of \ac{sAGB} stars which might otherwise be expected to undergo \ac{ECSN}.  In the \fiducial  variation, single stars with \ac{ZAMS} mass in the range ${\approx}\unit[7.5-8.1]{\Msun}$ yield \acp{ECSN}.  This constitutes ${\sim} 13\%$ of \ac{NS} progenitors, assuming a \citet{Kroupa01} initial mass function and a \ac{ZAMS} mass range extending to \unit[20]{\Msun} for \ac{NS} progenitors.  Fig.~\ref{fig:cdfs} and \ref{fig:pflow} suggest that a contribution of no more than a few per cent from non-interacting \ECps would be more consistent with observations.  This would require a reduction in the width of the \ac{ZAMS} mass range for \ac{ECSN} progenitors to ${\lesssim}\unit[0.2]{\Msun}$, or removing this possibility altogether, as in our \noWideECSNe and \noWideECSNeKickOne models. This is consistent with \citet{Doherty17}, who predict an \ac{ECSN} progenitor \ac{ZAMS} mass range of ${\approx}\unit[9.5-9.7]{\Msun}$ at \Zsun (and indeed a width of ${\approx}\unit[0.2]{\Msun}$ across all metallicities), 
as well as \citet{Tarumi:2021} who find a similarly narrow mass range for \ac{ECSN} progenitors from dwarf galaxy Sr abundances \citep{Hirai_strontium:2019}.  Our results are qualitatively similar at $\Zsun/10$. We do not distinguish between \acp{ECSN} and low-mass iron-core \acp{CCSN} in our models \citep{PodsiadlowskiEtAl:2004}; an increased yield from the latter, if they produce comparably low kicks, would then require a proportional reduction in the former.  We have not considered here the possibility that accretion-induced collapse of white dwarfs may lead to a weakly-kicked \acp{NS}, as the models for this channel remain very uncertain \citep{Nomoto:1991,RuiterEtAl:2019, WangLiu:2020}.

Although no \ac{ECSN} progenitors have been observationally confirmed, some candidate populations have been proposed which, if validated, would help to constrain the nature of \acp{ECSN} in single stars \citep{Ogrady20}.  Indeed, \citet{Hiramatsu21} propose that SN 2018zd was an \ac{ECSN} from a single \ac{sAGB} star based on the low energy and chemical profile of the light curve.  However, we find this to be unlikely if \acp{ECSN} are indeed very rare in single stars.  
Additionally, the light curves for stripped \acp{ECSN} may be very short, rendering detection especially challenging, which could explain the dearth of \ac{ECSN} candidates. A possible alternative to reducing the \ac{ZAMS} mass range for producing \acp{ECSN} from single stars would be reducing the observability of the remnant as a young radio pulsar.

We obtain an additional constraint on natal kicks by considering their effect on \ac{DNS} merger rates.  Galactic double neutron star and gravitational-wave observations indicate a local \ac{DNS} merger rate of $\unit[430^{+280}_{-130}]{\textrm{Gpc}^{-3} \textrm{yr}^{-1}}$ \citep{Pol:2020} and  $\unit[320^{+490}_{-240}]{\textrm{Gpc}^{-3} \textrm{yr}^{-1}}$  \citep{GWTC2:pop}, respectively.  Given a local star formation rate of $\unit[1.5 \times 10^7]{\Msun \textrm{Gpc}^{-3} \textrm{yr}^{-1}}$ \citep{MadauDickinson:2014}, this implies a yield of ${\sim} 25$ merging \acp{DNS} per $\unit[10^6]{\Msun}$ of star formation.  Our models predict a lower \ac{DNS} yield of 7--8 merging \acp{DNS} per $\unit[10^6]{\Msun}$ of star formation (see Table \ref{tab:stats}).  However, some of the locally merging \acp{DNS} formed at higher redshifts when the star formation rate was higher, so this yield is not inconsistent with observations.  Removing \acp{ECSN} in wide binaries does not reduce the merging \ac{DNS} yield since such systems are very unlikely to form \acp{DNS} merging within a Hubble time.

Our analysis is limited by the size of the observational data set and by the possibility of selection effects in choosing which pulsars are followed up with VLBI.  A complete VLBI follow-up within a predetermined volume could address both concerns.


\emph{Acknowledgements:} We thank Derek Bingham, Giulia Cinquegrana, Dan Foreman-Mackey, Jonathan Gair, Andrei Igoshev, Philipp Podsiadlowski, Ryan Shannon, Steinn Sigurdsson, and members of Team COMPAS for useful discussions.  The authors are supported by the Australian Research Council Centre of Excellence for Gravitational Wave Discovery (OzGrav), through project number CE170100004.  This work made use of the \mbox{OzSTAR} high performance computer at Swinburne University of Technology.  \mbox{OzSTAR} is funded by Swinburne University of Technology and the National Collaborative Research Infrastructure Strategy (NCRIS).  IM and ARD are recipients of the Australian Research Council Future Fellowships (FT190100574 and FT150100415, respectively).  AVG acknowledges support by the Danish National Research Foundation (DNRF132).
\software{Simulations in this paper made use of the COMPAS\footnote{COMPAS is freely available at \url{http://github.com/TeamCOMPAS/COMPAS}.} rapid binary population synthesis code, v02.19.02.  Data was taken (partially) from the ATNF Pulsar Catalogue \citep{psrcat}, and analysis was performed using python v3.6 \citep{python}, numpy v1.19 \citep{numpy}, matplotlib v3.3 \citep{matplotlib}, and astropy v4.1 \citep{astropy:2013, astropy:2018}.}


\bibliographystyle{aasjournal}
\bibliography{bib.bib}

\end{document}